\documentclass[10pt,aps,prl,onecolumn,nopacs,final,letterpaper,superscriptaddress,longbibliography]{revtex4-2}
\usepackage[utf8]{inputenc}
\usepackage{amsmath,amsfonts,amssymb}
\usepackage{graphicx}
\usepackage{array}
\usepackage{bm}
\usepackage{hyperref}
\usepackage{xcolor}
\usepackage{quotes}
\usepackage{float}
\definecolor{goodblue}{RGB}{0, 91, 187}
\hypersetup{
  colorlinks=true,
  allcolors=goodblue,
  urlcolor=goodblue,
  citecolor=goodblue,
  pdfborder={0 0 0},
  breaklinks=true,
}

\begin{document}

\title{Winning Opinion in the Voter Model: Following Your Friends' Advice or That of Their Friends?}
\author{Francisco  J. Mu\~noz}
\email{francisco.munoz@urjc.es}
\affiliation{Departamento de Matem\'atica Aplicada, Ciencia e Ingenier\'ia de Materiales y Tecnolog\'ia Electr\'onica, Escuela Superior de Ciencias Experimentales y Tecnología, Universidad Rey Juan Carlos, 28933, M\'ostoles (Madrid), Spain}

\author{Juan Carlos Nu\~no}
\email{juancarlos.nuno@upm.es}
\affiliation{Departamento de Matem\' atica Aplicada, Universidad Polit\'ecnica de Madrid,  28040-Madrid, Spain}

\begin{abstract}

We investigate a variation of the classical voter model in which the set of influencing agents depends on an individual's current opinion. The initial population consists of a random sample of equally sized sub-populations for each state, and two types of interactions are considered: (i) direct neighbors, and (ii) second neighbors (friends of direct neighbors, excluding the direct neighbors themselves). The neighborhood size, reflecting regular network connectivity, is kept constant across all agents. Our findings reveal that varying the interaction range introduces asymmetries that influence the probability of consensus and convergence time. At low connectivity, direct neighbor interactions dominate, driving consensus. As connectivity increases, the probability of consensus for either state becomes equal, mirroring symmetric dynamics. It is shown that this asymmetric effect on the probability of consensus is independent of the network topology in small-world and scale-free networks. Asymmetry is also reflected in convergence time: while symmetric cases show decreasing times with increased connectivity, asymmetric cases exhibit an almost linear increase. Contrary to the probability of achieving consensus, the effect of asymmetry on the consensus time does depend on the network topology. The introduction of stubborn agents further accentuates these effects, particularly when they favor the less dominant state, significantly increasing consensus time. We conclude by discussing the implications of these findings for decision-making processes and political campaigns in human populations.

\end{abstract}

\maketitle

\section{Introduction}

Human behavior, especially the formation and evolution of opinions, is influenced by a complex interplay of factors, many of which remain poorly understood \cite{Galam, Sen, Castellano}. Among these factors, social ties play a crucial role in shaping individual opinions and preferences.

Humans inherently form networks through which information is exchanged. These networks can vary significantly depending on the context. For instance, connections formed among university peers may differ from those formed in romantic or social contexts. Nonetheless, individuals often participate in multiple overlapping networks \cite{Kivela, Amato}. 

The way an individual's ``neighborhood of influence'' is defined can significantly affect opinion dynamics. Even when network structures are uniform, the interaction rules can determine which opinions prevail (see, for example, \cite{Avella, Tessone,Schulze}). This study explores whether individuals benefit more by following the opinions of their immediate friends or their friends' friends.

The question was inspired by a personal observation: ``When I evaluate the potential success of a publication, I find that my friends' feedback often lacks objectivity. In contrast, their friends' assessments, though less personal, tend to be more accurate and unbiased.'' Motivated by this, we investigate the dynamics of opinion formation under two scenarios: when agents are influenced by direct friends or by friends of friends.

In this study, we simulate opinion dynamics in a population of $N$ agents (nodes) on a regular ring network with circular boundary conditions. Each agent can adopt one of two states (e.g., 0 or 1). Agents in state 0 are influenced by their direct neighbors, while those in state 1 are influenced by their second neighbors (excluding direct neighbors). Using the classical voter model \cite{Clifford, Holley, SanMiguel}, with constant size and asynchronous updates, we analyze the probability of consensus and the time required to achieve it under symmetric and asymmetric neighborhood definitions. 

The influence of network topology is investigated using two models: the Watts-Strogatz small-world network \cite{WS1998} and the Barab\'asi-Albert scale-free network \cite{BA1999}. In the Watts-Strogatz model, a rewiring mechanism transforms a regular cyclic network into a random network by controlling the probability of relinking nodes, denoted by $p$. In contrast, the Barab\'asi-Albert model generates a heterogeneous network via preferential attachment, which results in a power-law distribution of connectivity. Our results indicate that while the effect of asymmetry in the neighborhood of influence is independent of the network structure, the topology does influence the time required for the population to achieve consensus, not by altering its qualitative shape but rather in a quantitative manner.

\begin{figure}[H]
\centering
\includegraphics[width=0.6\textwidth]{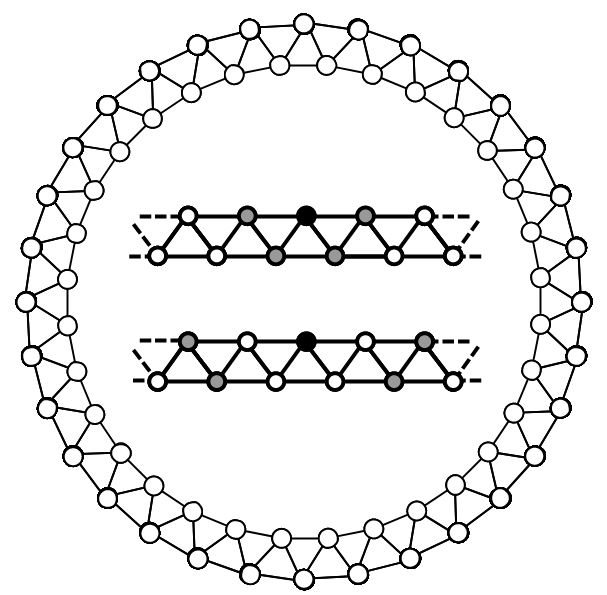}
\caption{Opinion dynamics occur on a regular network as depicted in this figure. The central inset shows the different neighborhoods that are considered in this model: (Top) Nearest neighbors and (Bottom) Second neighbors (excluding the common neighbors). In this example, the network size is: $N=64$ and the connectivity is: $k=4$ (for both neighborhoods). The maximum connectivity that assures no intersection between both neighborhoods is $k_{max}=30$.}
\label{fig1}
\end{figure}

\section{Voter model on a regular cyclic network}\label{Voter}

We consider a population of $N$ agents, each adopting one of two states (0 or 1), arranged on a regular cyclic network with connectivity $k$. Initially, the population is evenly divided between states 0 and 1, with agents randomly distributed across the network. Simulations are carried out until a final time of 1 million, guaranteeing that the population reaches one of the two consensus state, either 0 or 1. Note that the maximum connectivity that assures that the two neighborhoods First-First and First-Second have no common agents is $k_{max}=\frac{N}{2}-2$.

According to the classical voter model \cite{Gracia}, at each time step, a node is randomly selected, and its state is compared to that of another randomly chosen node within its neighborhood. If the two states differ, the first node adopts the second node's state (see Fig. \ref{fig1}). One time step is defined as $N$ updates, ensuring that, on average, each node is updated once.

In the symmetric case, neighborhoods are identical for both states ($k$-nearest neighbors). In the asymmetric case, neighborhoods depend on the agent's state: state 0 agents interact with nearest neighbors, while state 1 agents interact with their second neighbors (avoiding the nearest neighbors of the central node). We analyze the probability of reaching a consensus (all agents in one state) and the time required for this consensus under both scenarios.

In the following subsections, we present the results regarding: (i) the probability of reaching a consensus at either 0 or 1 and (ii) the time required for the population to achieve such a consensus. Then, in Section \ref{stubb}, we analyze these measures in the presence of a stubborn agent (also known as a zealot), and in Section \ref{topo}, we explore the influence of network topology.

\begin{figure}[H]
\centering
\includegraphics[width=0.7\textwidth]{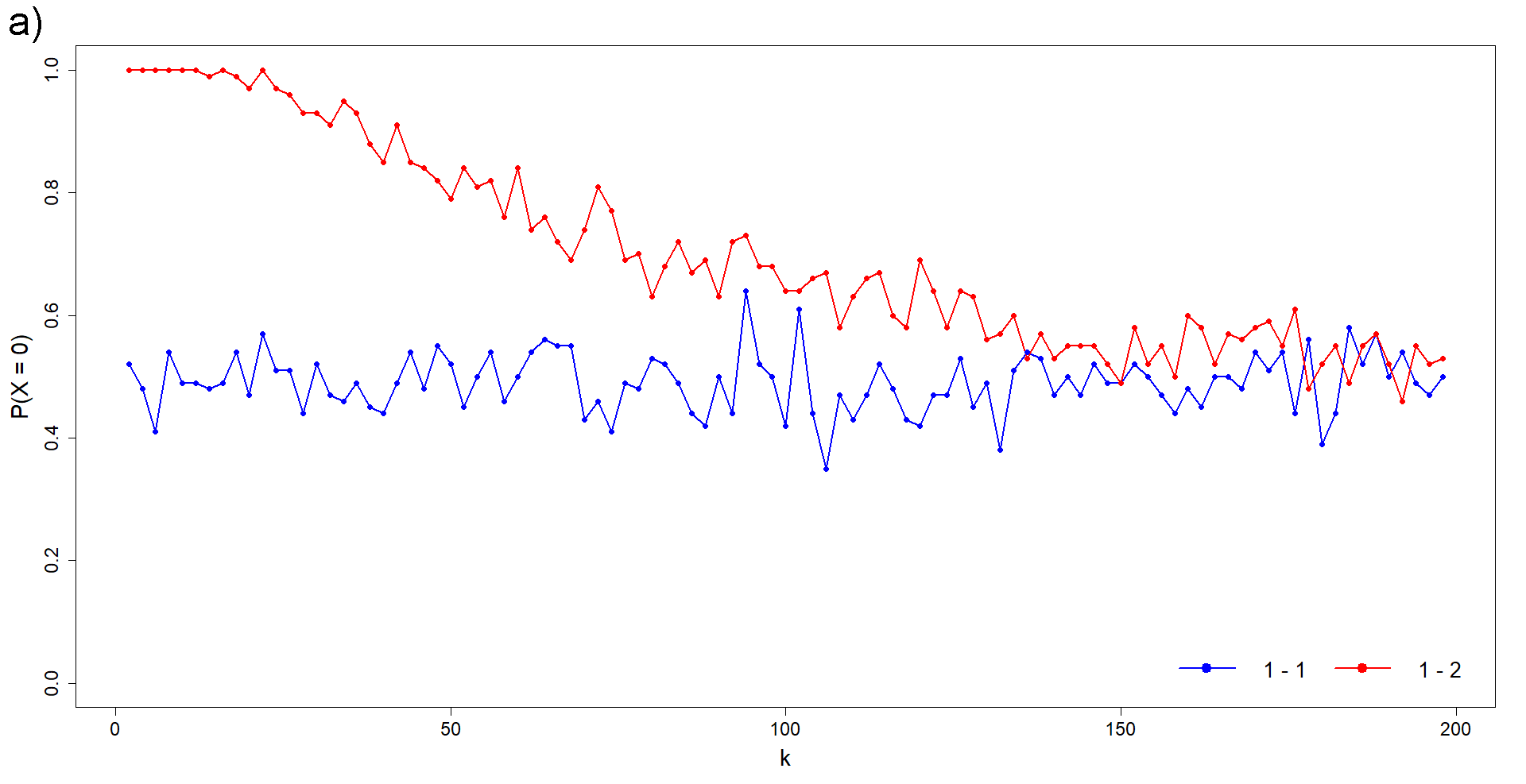}
\includegraphics[width=0.7\textwidth]{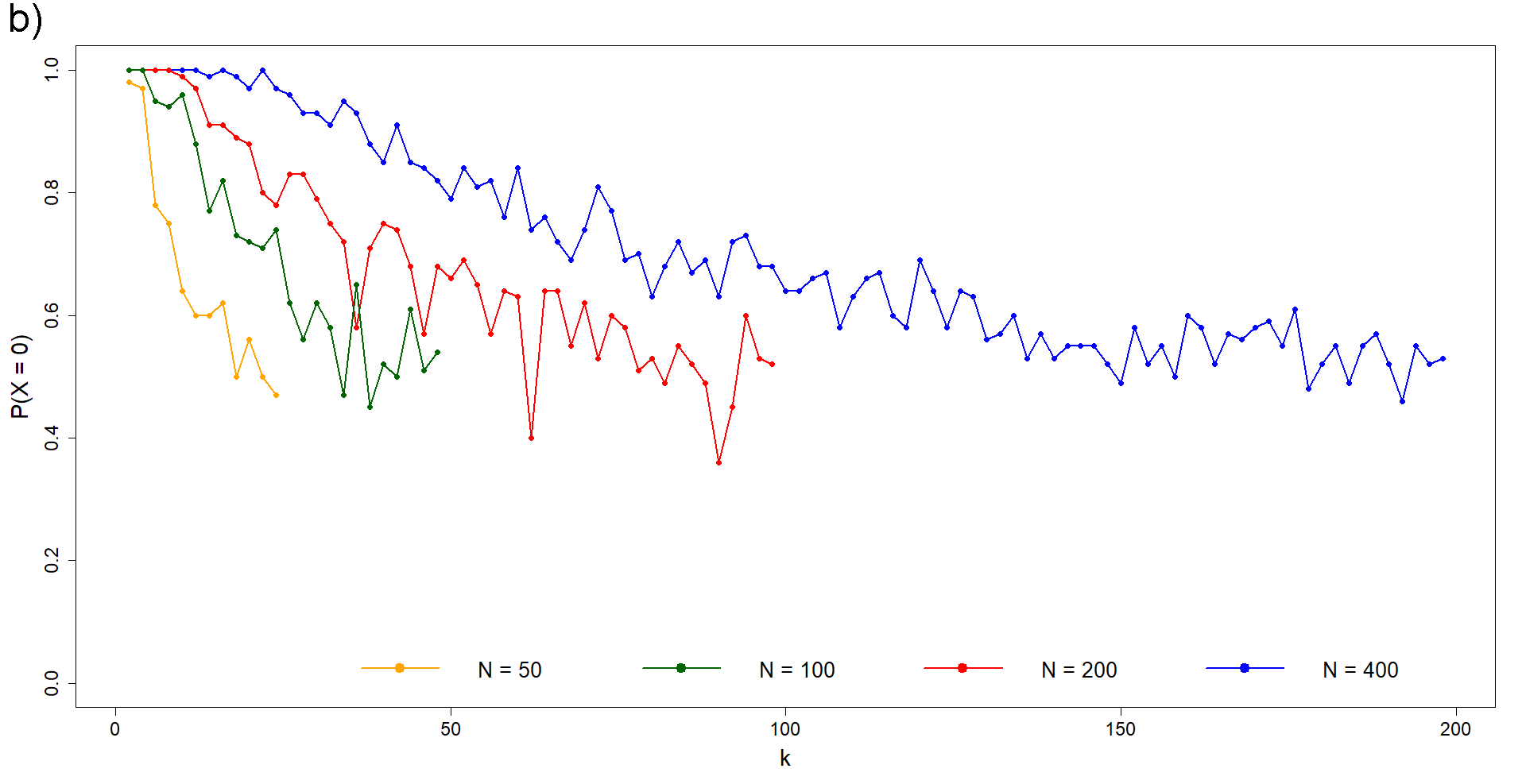}
\includegraphics[width=0.7\textwidth]{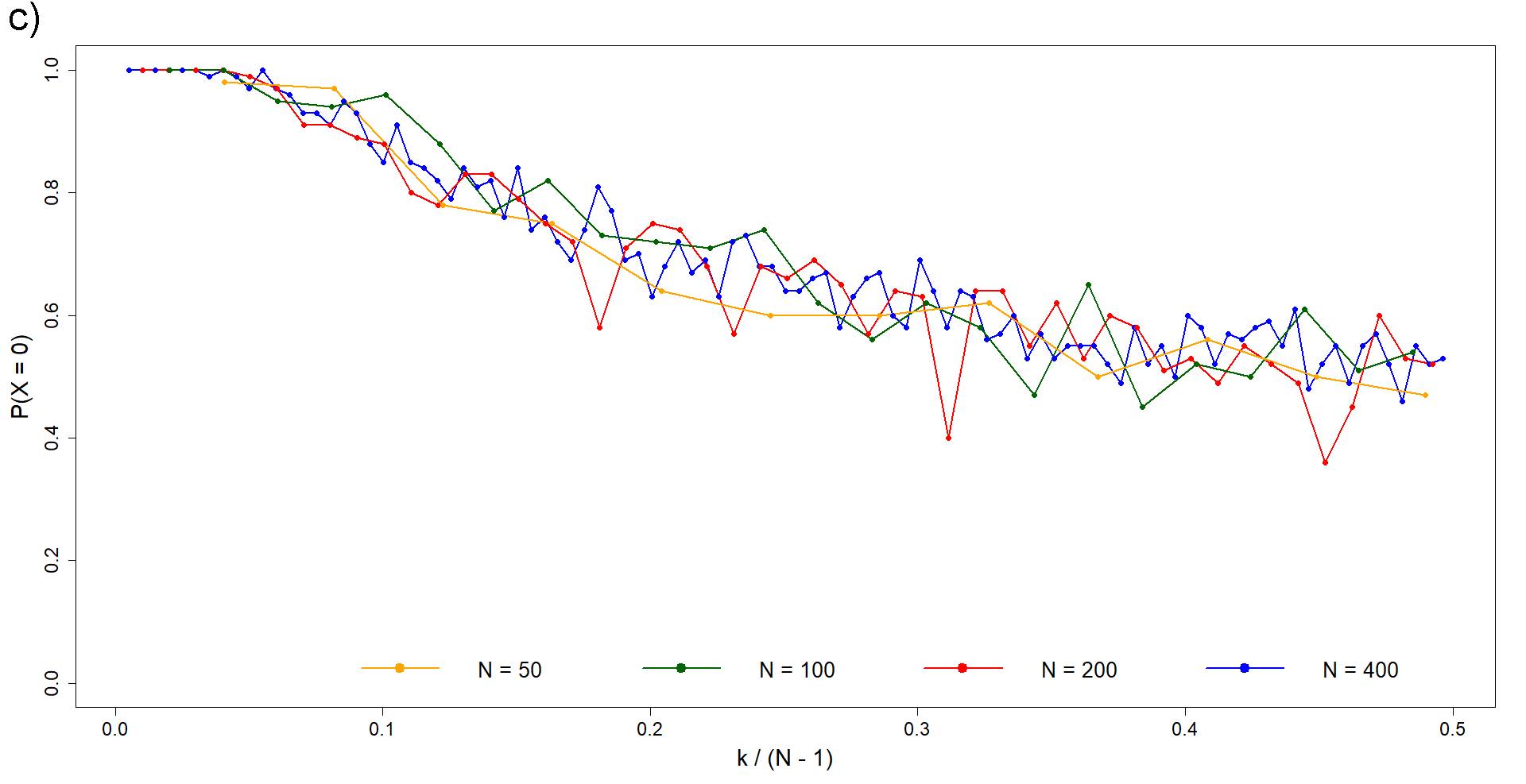}
\caption{(a) Probability of achieving a consensus population of individuals at state 0, $P(X=0)$, as a function of the network connectivity $k$ for the two cases: (blue) symmetric population, i.e. both states have the same nearest neighborhood (First-First), and (red) asymmetric population where individuals at state 0 interact with their nearest neighbors, while individuals at state 1 interact with their second neighbors (First-Second, avoiding first neighbors). In contrast to the symmetric case, where this probability is independent of $k$, when each state has a different neighborhood, $P(X=0)=1$ for low values of the connectivity and decreases monotonously to 1/2 as $k$ increases, approaching the symmetric dynamics. The probability is estimated from 100 simulations, and the network size is $N=400$. 
 (b) The same probability $P(X=0)$ in terms of $k$ for four networks sizes (estimated from the same number of simulations). The maximum connectivity that assures that the two neighborhoods do not intercept is $k_{max}=\frac{N}{2} -2$. (c) As the above pannel but normalizing the $X$-axis by the maximum connectivity $N-1$. As seen, the four curves collapse in one, showing the asymptotic decrease to the symmetric probability $P(X=0)=1/2$ as $k$ increases.}
\label{fig2}
\end{figure}

\subsection{Probability of reaching a consensus population}

For symmetric neighborhoods, the probability of reaching a consensus state (0 or 1) is equal (50$\%$) and independent of connectivity $k$ \cite{DeGroot,Sen}. However, in the asymmetric case, the probability depends on $k$ (see Fig. \ref{fig2}a for the probability of achieving consensus at state 0). At low $k$, consensus strongly favors state 0 (with probability 1). As $k$ increases, the probability of consensus for state 1 rises, eventually reaching parity with state 0. 

To understand the origin of this asymmetry, we analyze the system’s dynamics over the course of a few steps. Remarkably, after a single time step, the probabilities of reaching either state, $x = 0$ (local neighborhood) or $x = 1$ (nonlocal neighborhood), are equal. The asymmetry—arising from the different neighborhoods associated with each state—emerges after two steps. We focus on the case $k = 2$ and examine the smallest interacting set that can influence the state of the central agent after two steps. In this case, this set consists of 9 nodes. We compute the probabilities of the system reaching either state, $x = 0$ or $x = 1$, after two time steps, considering all $2^9 = 512$ possible equiprobable initial configurations.

The following updating equation allows us to compute the state of any node $x_i$, for $i=-2,\ldots, 2$, at step $n+1$ as a function of the states of its neighbors, dependent on its own state, at the previous step $n$:
\begin{eqnarray}\label{recursive1}
x_i(n+1) &  =  & F(x_{i-2}(n), x_{i-1}(n),x_i(n), x_{i+1}(n), x_{i+2}(n))
\end{eqnarray}
for $i=-2,\ldots,2$ and $F$ is given by:
\begin{eqnarray}\label{recursive2}
F(x_{i-2}(n), x_{i-1}(n), x_i(n), x_{i+1}(n), x_{i+2}(n)) & = &  (1-x_i(n)) (x_{i-1}(n)\,x_{i+1}(n)  \nonumber \\
&   + & \mathcal{B}(0.5)\, (x_{i-1}(n)+x_{i+1}(n)  \nonumber \\
&  -  & 2\,x_{i-1}(n)\,x_{i+1}(n)))  \nonumber \\
&   + &   x_i(n) (x_{i-2}(n)\,x_{i+2}(n)  \\
&   + & \mathcal{B}(0.5)\, (x_{i-2}(n)+x_{i+2}(n) \nonumber \\
& -  & 2\,x_{i-2}(n)\,x_{i+2}(n))) \nonumber
\end{eqnarray}
Here $x_{i-1}$ and $x_{i+1}$ and $x_{i-2}$ and $x_{i+2}$ denote the states of the left and right first and second neighbors for each node $i$ and $\mathcal{B}(0.5)$ represents a Bernoulli distribution of $p=1/2$ in $\{0,1\}$. Note that the probabilistic terms arise due to the tie breaking rule, applied when an equal number of neighbors are in each state. \footnote{It is worthy to remark that other probabilities for the tie breaking rule yields different results (work in progress).}

The update equation for the state of the central node at time $n+2$ is:
\begin{eqnarray}\label{recursive3}
x_0(n+2) &  =  & F(x_{-2}(n+1), x_{-1}(n+1),x_0(n+1), x_{+1}(n+1), x_{+2}(n+1))
\end{eqnarray}

For each initial configuration of the nine sites, we obtain the probability of the central site being in state $x=0$ after two steps. The results show that this probability exceeds 0.5, specifically around 0.515625. This clearly contrasts with the symmetric case, where the probability is exactly 0.5.

This asymmetry explains why opinions based on local neighborhoods eventually dominate the entire population. A similar recurrent equation, although considerably longer, can be derived for the case $k=4$, being the probability of the central site in state $x=0$ after two steps estimated as 0.508174 (see Appendix \ref{apend1}).

The probability of reaching the 0-state consensus depends on the network size $N$ (Fig. \ref{fig2}b). However, when plotted against normalized connectivity $\frac{k}{N-1}$, these probabilities collapse onto a single curve across different population sizes, indicating a universal dependence on the normalized parameter, as shown in Fig. \ref{fig2}c.

\begin{figure}[H]
\centering
\includegraphics[width=0.8\textwidth]{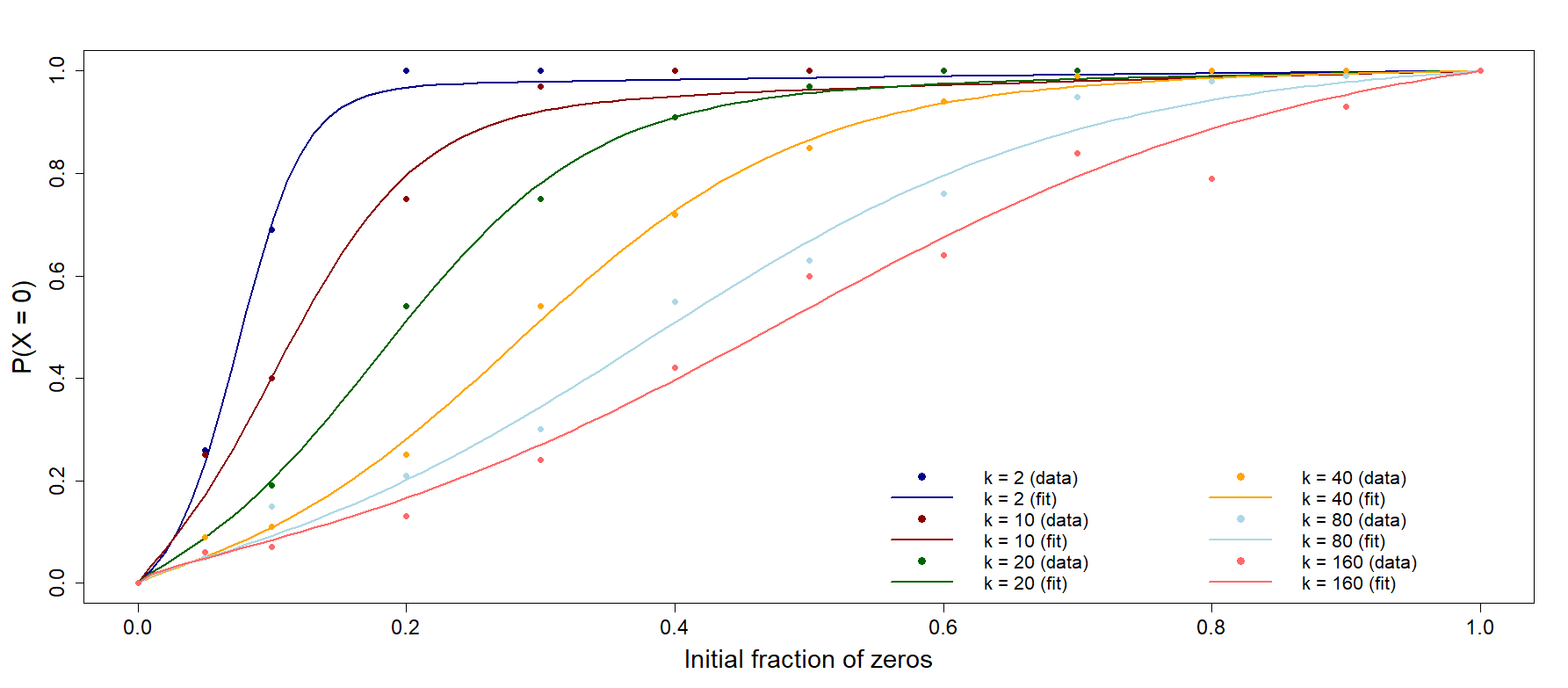}
\caption{The probability of achieving 0-consensus in the population as a function of the initial fraction of agents in the 0-state, randomly distributed across the network, is shown for several values of $k$ in the asymmetric case. Each data point is obtained from 100 simulations for a network with size $N=400$. For each value of $k$, the points can be well fitted to the function (\ref{ffit}) defined in the main text, using the parameters described in Table \ref{tab:modelo_k}.}
\label{fig2b}
\end{figure}

The probability of achieving 0-consensus in the population depends on the initial conditions. As stated, the results presented above are obtained using an equally distributed initial population of agents in states 0 and 1, i.e., with a $50\%$ fraction in state 0. Figure \ref{fig2b} shows the probability of achieving 0-consensus, $P(X=0)$, as a function of the fraction of the initial population in state 0, $\rho$. In all cases, the initial population is randomly distributed across the vertices of the network. For each value of $k$, these probabilities can be fitted to a family of functions that pass through the points $(0,0)$ and $(1,1)$:
\begin{equation}\label{ffit}
P(X=0) =\frac{1}{1 + e^{-B \,(\rho - C)}} - \frac{1}{1 + e^{B \, C}} - \sqrt{\rho} \,  \left(
\frac{1}{1 + e^{-B \, (1 - C)}} - \frac{2 + e^{B \, C}}{1 + e^{B \, C}}\right)
\end{equation}
see table \ref{tab:modelo_k} for the values of parameters $B$ and $C$ corresponding to each $k$.

\begin{table}[H]
    \centering
    \begin{tabular}{|c|c|c|c|}
        \hline
        $k$ & B & C & RSS \\
        \hline
        2   & 40.4144  & 0.07448  & 0.003055  \\
        10  & 19.0588  & 0.1025   & 0.01575   \\
        20  & 12.7847  & 0.1804   & 0.003038  \\
        40  & 9.3353   & 0.2819   & 0.004153  \\
        80  & 6.3784   & 0.3837   & 0.01538   \\
        160 & 5.3150   & 0.4897   & 0.02035   \\
        \hline
    \end{tabular}
    \caption{Fitted values for the parameters $B$ and $C$ of function (\ref{ffit}) for different $k$. Note the low values of the Residual Sum-of-Squares (RSS) in all cases.}
    \label{tab:modelo_k}
\end{table}

\subsection{Consensus time}

\begin{figure}[H]
\centering
\includegraphics[width=0.8\textwidth]{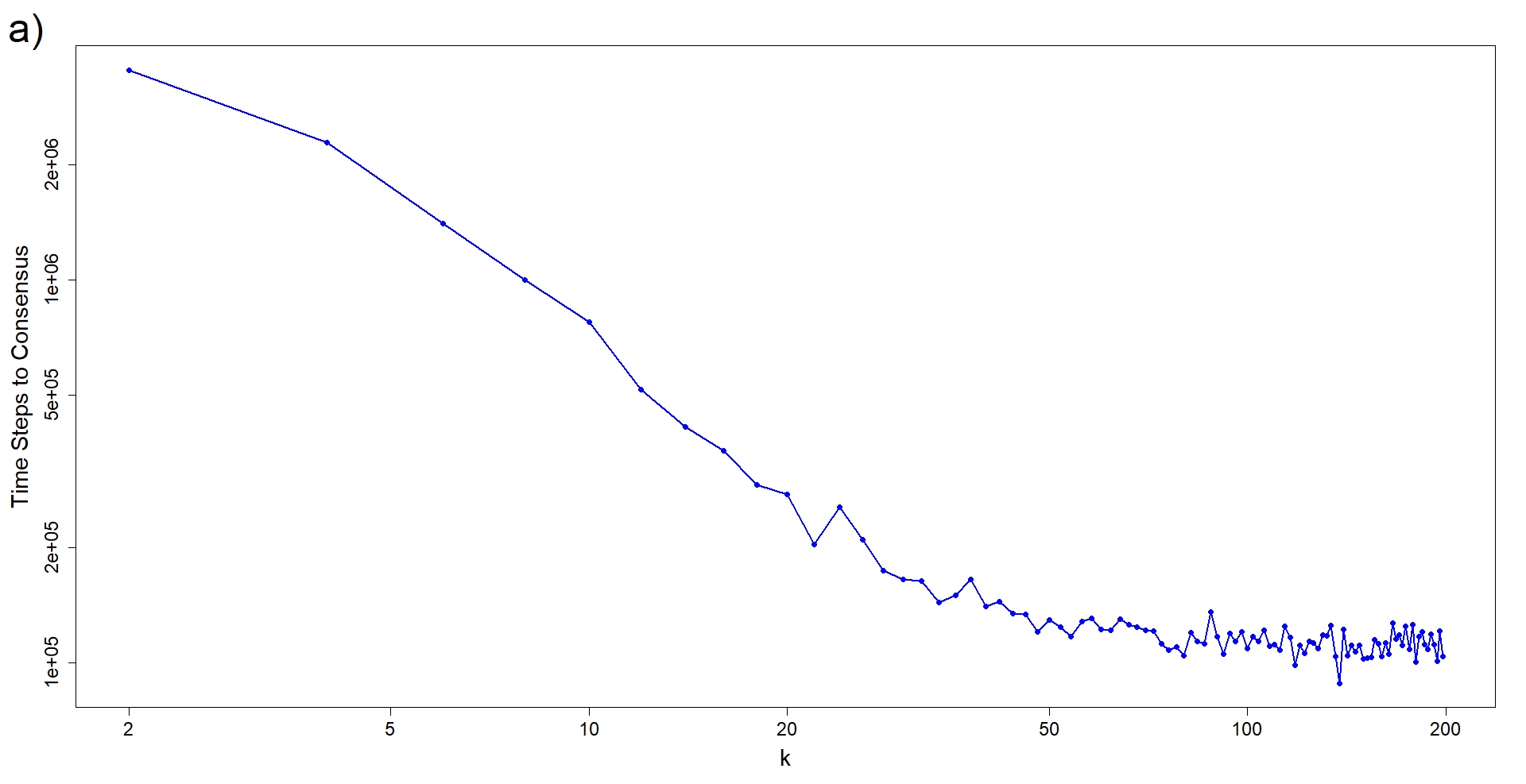}
\includegraphics[width=0.8\textwidth]{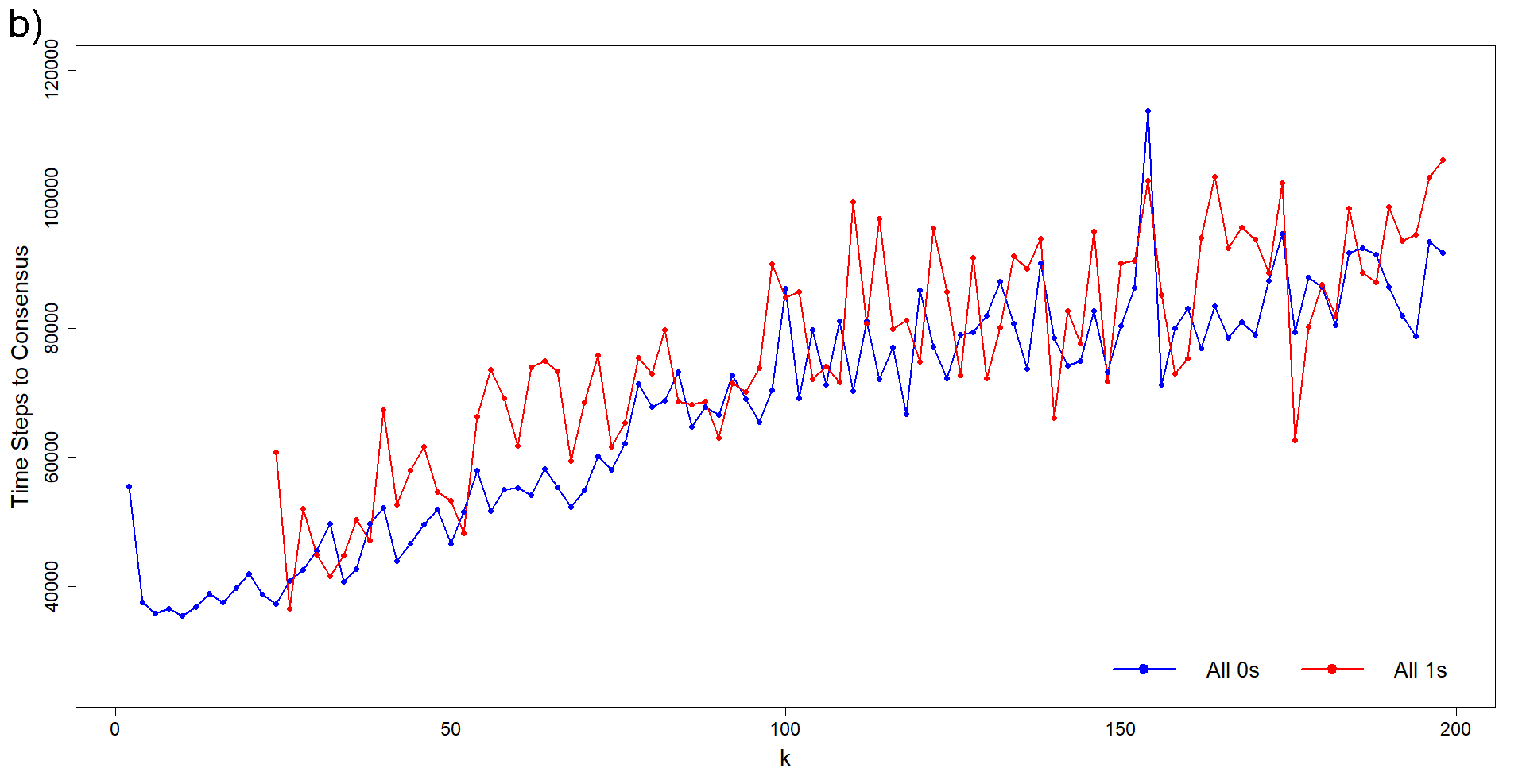}
\caption{Time steps to a consensus population as a function of the network connectivity $k$. Asynchronous updating on a network formed by 400 nodes (agents). (a) Symmetric case: both states have the same nearest neighborhood, (b) Asymmetric case: nodes at state 0 interact with their nearest neighbors and nodes at state 1 interact with their second neighbors, avoiding the common neighbors. Blue curve corresponds to the 0-state consensus and red curve to 1-state consensus. Note that for low $k$-values only red points exist since reaching the 0-state consensus has probability 1. Importantly, the dependence of the consensus time with $k$ differs drastically between the two figures, while it decreases almost exponentially (a straight line for intermediate values of $k$ in a semilogarithmic scale) for the symmetric case, it increases (almost linearly) with $k$ for the asymmetric case. Moreover, as it can be observed, the consensus time in the asymmetric case is lower than in the symmetric one.}
\label{fig3}
\end{figure}

Figure \ref{fig3} shows the results obtained from the simulations described in the previous section, where the consensus time for each state was recorded. Consensus time is defined as the first step at which the entire population adopts a single state (either 0 or 1), and it is calculated separately for each state. The reported consensus times represent the average of all simulations that result in a particular state.

For large connectivity values, the consensus times for both states are similar. However, for low connectivity values in the asymmetric case, consensus is only achieved for state 0. As shown in Figure \ref{fig3}a, for the symmetric case (where both states have identical neighborhoods), the consensus time decreases with increasing network connectivity, following a quasi-exponential trend. In contrast, for the asymmetric case, the consensus time increases with connectivity, as illustrated in Figure \ref{fig3}b.

\begin{figure}[H]
\centering
\includegraphics[width=0.8\textwidth]{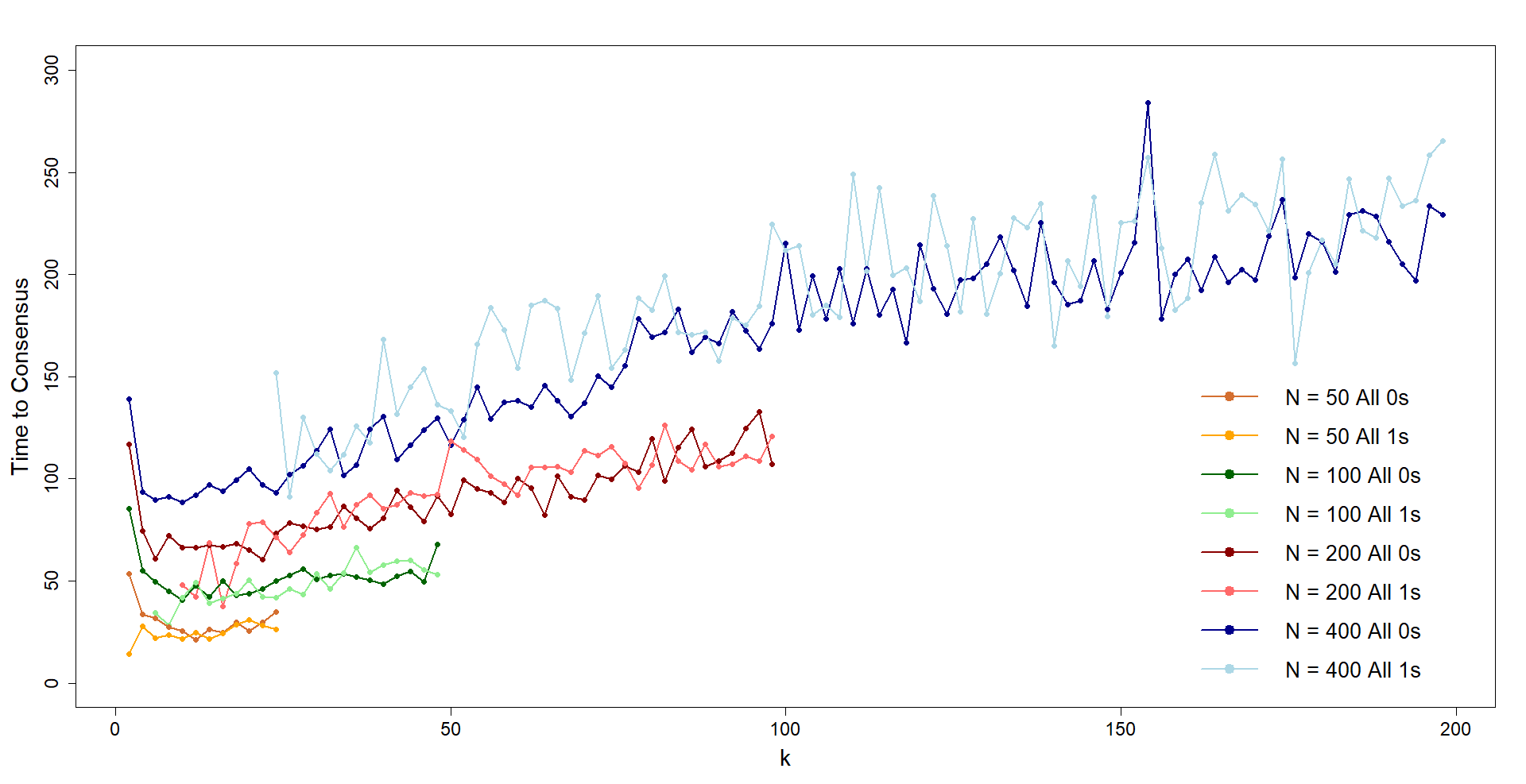}
\caption{Time until consensus, i.e. time steps divided by $N$, for an initial population to reach a consensus, as a function of the network connectivity $k$ for the asymmetric cases for different network sizes: 400 (blue), 200 (red), 100 (green) and 50 (orange). Light colors correspond to the time to 1-state consensus and dark colors to 0-state consensus. Note that for low $k$-values for the four $N$ sizes, only dark points appear since only consensus around the 0-state occurs. For this state, as it can be observed, the consensus time exhibits a minimum value for the four sizes, at a connectivity (normalized) value that depends on $N$. The four curves can be fitted to a straight line: $\tau = m \, k + n$,  with the same slope $m$ and the intercept $n=f(N)$. The maximum connectivity that assures that the two neighborhoods do not intercept is $k_{max}=\frac{N}{2} -2$.}
\label{fig4}
\end{figure}

The behavior shown in Fig. \ref{fig3}b for the asymmetric case, with a network size of $N=400$, is consistent across other network sizes (see Fig. \ref{fig4}). For low $k$ values, where consensus exclusively occurs in the 0-state population, the functions representing the time to consensus (measured as time steps divided by network size) increase linearly with the normalized network connectivity, $\frac{k}{N-1}$. Notably, the slope of this linear increase becomes steeper for larger network sizes, $N$.

The discrepancy between the symmetric and asymmetric cases regarding how consensus time depends on connectivity is not entirely understood. However, it can be attributed to competition between the different neighborhood structures, which creates an asymmetry in the probabilities of reaching consensus. Specifically, the probability of achieving consensus in state 0 is higher than that for state 1 (see Fig. \ref{fig2}a). This asymmetry in probabilities results in a shorter time to reach consensus, particularly at low $k$-values.

As connectivity ($k$) increases, the probabilities of reaching consensus for states 0 and 1 converge, as does the time required to achieve consensus (Fig. \ref{fig3}b). It is also noteworthy that for $k>10$, the consensus times for both states are effectively equal (see the red and blue curves in Fig. \ref{fig3}b). For $k<10$, where the probability of reaching consensus in state 0 is nearly 1, the consensus time increases. This behavior is similar to the symmetric case, where both states have identical neighborhoods.

\begin{figure}[H]
\centering
\includegraphics[width=0.8\textwidth]{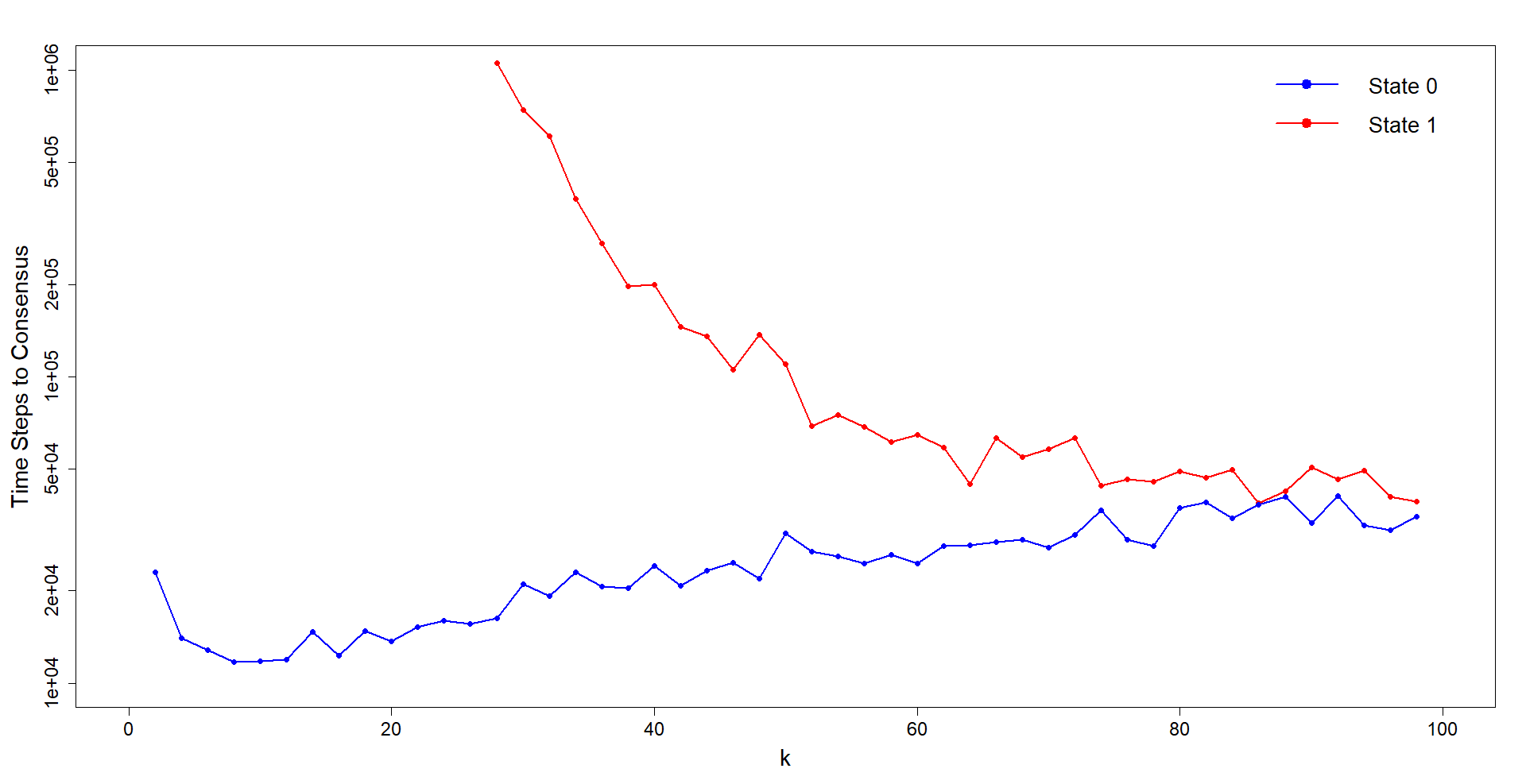}
\caption{Time to consensus for both states in the presence of a stubborn agent (node), either in state 0 (blue) or 1 (red) in a network with $N=200$ nodes. Note that for $k<28$ the consensus times when the state of the stubborn agent is 1 is not depicted due to its extremely large values (larger than $10^7$ time steps). As in the asymmetric case without stubborn agents, the time to 1-consensus is larger than the corresponding time for the 0-consensus for all $k$-values, although these times converge as $k$ increases.}
\label{fig5}
\end{figure}

\section{Effects of stubborn agents}\label{stubb}

To better understand the effect of neighborhood asymmetry on the opinion dynamics of the population, we consider the inclusion of agents whose opinions remain fixed over time, known as stubborn agents. It is well established that the presence of stubborn agents supporting both states can prevent the population from achieving consensus \cite{Mobilia2007}.

In the simplest case of a single stubborn agent, the dynamics of the classical voter model with nearest neighbors are biased toward the agent's state. For instance, if the stubborn agent permanently holds the 0-state, the entire population will eventually converge to this state, regardless of its initial configuration \cite{Mobilia2003}. In essence, the stubborn agent's state dominates and overrides the other state. The time required to achieve this consensus can also be determined in the infinite population limit \cite{Mobilia2003}.

In asymmetric neighborhoods, the effect of a stubborn agent remains similar to that in the symmetric case: the agent drives the population to consensus in its own state. However, the time required to reach this consensus is significantly affected by the presence of the stubborn agent (see Fig. \ref{fig5}). If the stubborn agent holds the 0-state (nearest neighbor interactions), the consensus time is significantly reduced for any value of $k$. Conversely, if the stubborn agent holds the 1-state (second-neighbor interactions), the consensus time increases dramatically for low $k$-values. In this case, the stubborn agent effectively neutralizes the natural tendency of the population to converge to the 0-state.

\section{Influence of network topology}\label{topo}

To isolate the effect of asymmetric neighborhoods on population dynamics, we begin by considering a regular cyclic network where each node has exactly $k$ first neighbors. This setup ensures that local inhomogeneities do not contribute to the asymmetric effects of neighborhoods on consensus formation. In this section, we examine how network topology influences opinion dynamics. To do so, we extend the previously defined cyclic network following the model introduced by Watts and Strogatz (1998) \cite{WS1998}. In this framework, the regular cyclic network corresponds to a rewiring probability of zero, while increasing the rewiring probability results in random reassignment of connections, altering key structural properties such as clustering and characteristic path length (see Fig. 2 in \cite{WS1998}). This raises the question of whether asymmetric opinion competition affects consensus formation and if so, how.

\begin{figure}[H]
\centering
\includegraphics[width=0.85\textwidth]{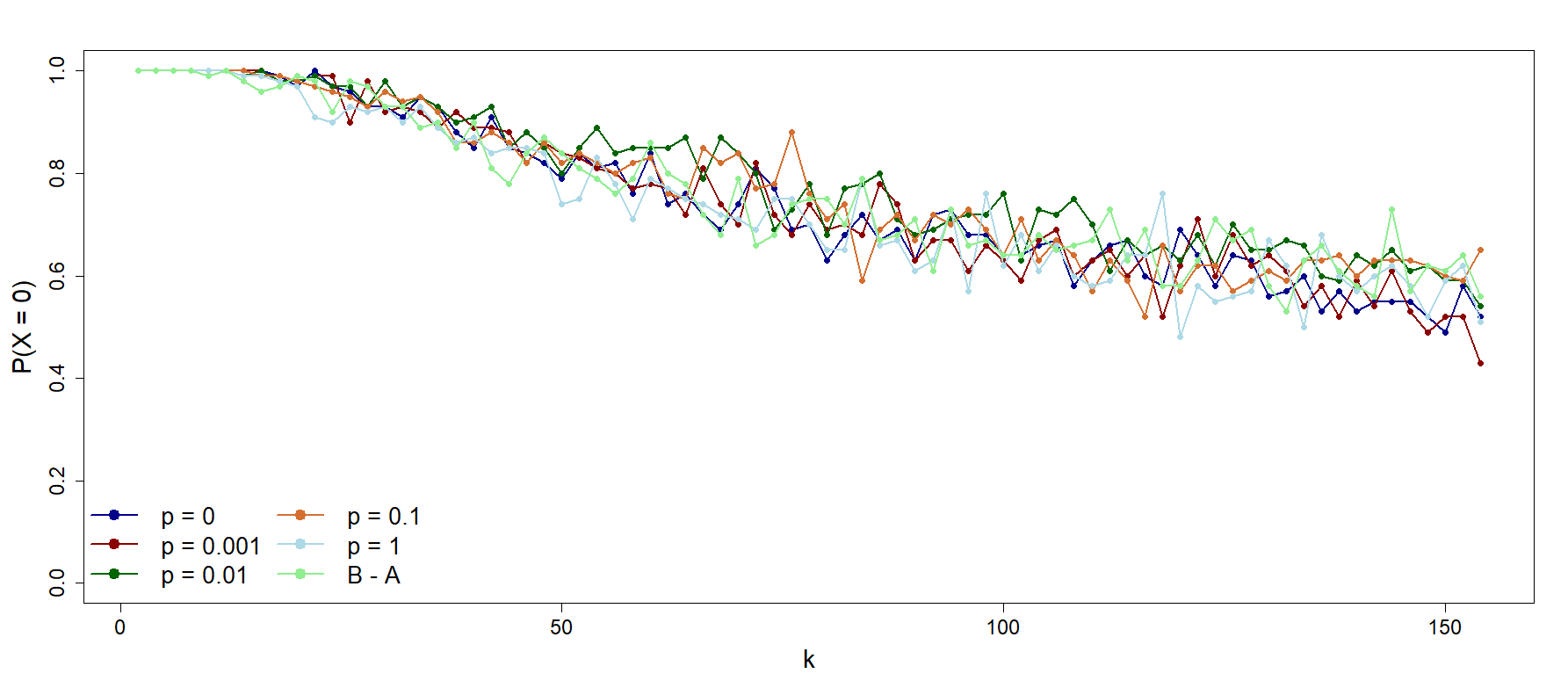}
\caption{Probability of achieving a 0-consensus population as a function of the average connectivity $k$ for various rewiring probabilities $p$ in a cyclic Watts-Strogatz network and for the B-A scale-free network. The initial population, evenly split between states 0 and 1, is randomly distributed along $N=400$ sites. Notably, the consensus probability appears independent of $p$; that is, the network's structure does not affect the final consensus state.
}
\label{fig6}
\end{figure}

To address this, we replicate the simulations from Section \ref{Voter} for a discrete set of rewiring probabilities, $p \in [0,1]$. The two extremes correspond to a regular cyclic network ($p=0$) and a fully random network ($p=1$). Figure \ref{fig6} depicts the probability of achieving a 0-consensus population (i.e., a steady state where all agents adopt state 0) in an asymmetric network, where state 0 agents interact with their nearest neighbors, while state 1 agents interact with their second neighbors, avoiding the nearest ones. The key finding is that, in this type of cyclic network, randomness in connectivity does not affect the probability of reaching a 0-consensus. In other words, neighborhood asymmetry emerges for low connectivity values irrespective of the rewiring probability $p$.

\begin{figure}[H]
\centering
\includegraphics[width=0.85\textwidth]{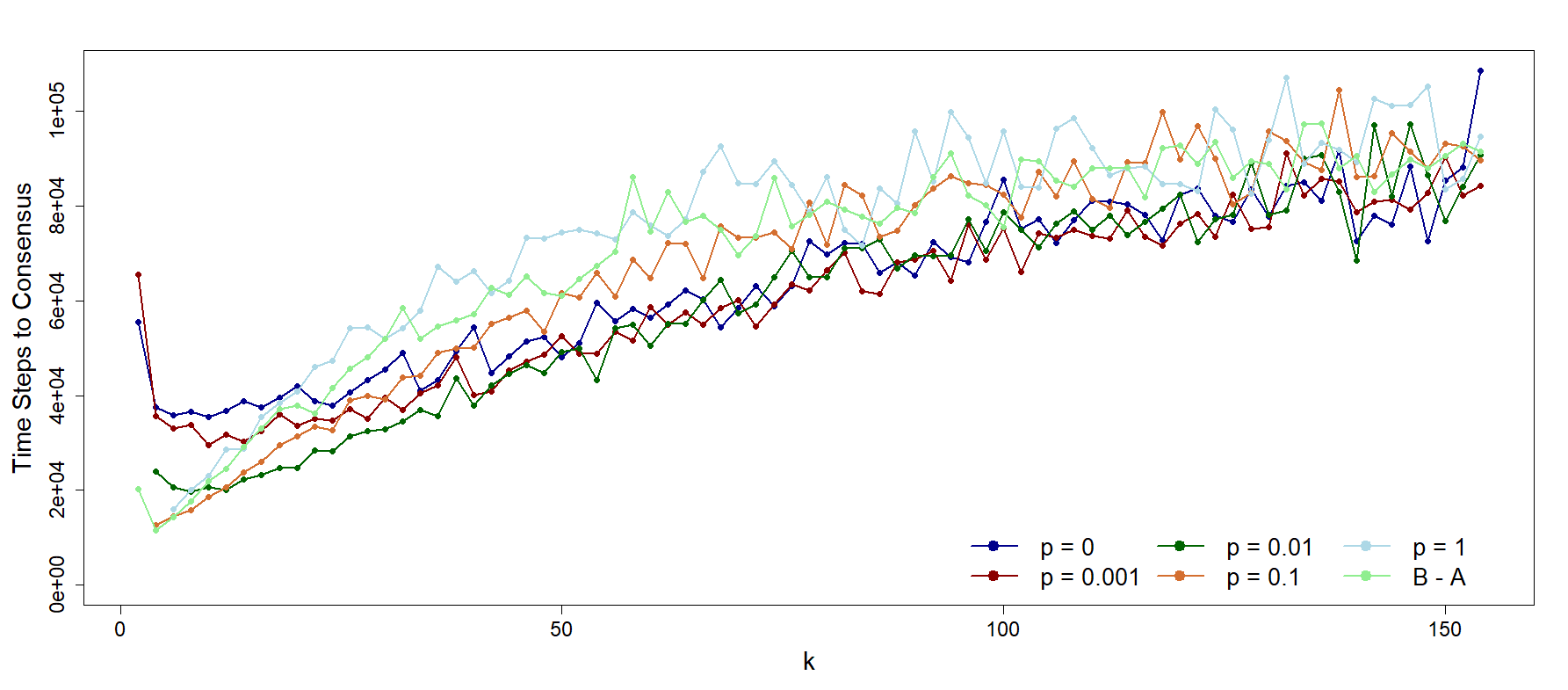}
\caption{Under the same conditions as the previous figure, we show the number of time steps to reach consensus as a function of the average connectivity $k$ for various $p$ values. While the overall trend increases linearly with $k$, longer times to achieve a 0-consensus population are observed for higher $p$ values and for the B-A scale-free network. In other words, more regular networks converge faster than either random ($p=1$) or inhomogeneous networks (B-A).}
\label{fig7}
\end{figure}

However, the time required to reach consensus does depend on $p$, as shown in Fig. \ref{fig7}. Notably, for $k>10$, consensus time increases linearly with $k$ for all values of $p$, with random networks ($p=1$) requiring more time to reach consensus compared to regular networks ($p=0$).

These results confirm that asymmetry between agent neighborhoods persists even when local neighborhoods vary, provided the network maintains a defined mean connectivity. To explore whether networks without a characteristic scale similarly influence opinion dynamics under asymmetric neighborhoods, we simulate the dynamics in a classical scale-free Barab\'asi-Albert (B-A) network \cite{BA1999}. As shown in Fig. \ref{fig6}, asymmetry in opinion dynamics remains evident even in this case: agents influenced by their first neighbors reach consensus more easily than those updating their opinions based on second neighbors, particularly at low connectivity values. However, as noted earlier, network topology does not significantly affect consensus probability when compared to cyclic Watts-Strogatz (W-S) Small-World networks with varying rewiring probabilities. That said, the time required to reach consensus does depend on topology. As illustrated in Fig. \ref{fig7}, the consensus time in the B-A network closely resembles that of a W-S Small-World network with $p=1$ (the fully random case).

\section{Discussion}

We study a population in which individuals interact on a circular network according to a classical voter model, where the state of each node determines its interaction neighborhood. Specifically, nodes in state 0 interact with their $k$ nearest neighbors, while nodes in state 1 interact only with the $k$ neighbors of each of their $k$ first neighbors (excluding common first neighbors). We show that, in contrast to the symmetric case, a randomly initialized population with an equal number of nodes in each state tends to evolve asymptotically toward a majority of nodes in state 0—that is, individuals who interact with their immediate neighbors. Furthermore, we demonstrate that the probability of reaching a consensus in which all nodes are in state 0 depends on both the connectivity $k$ and the network size $N$. In the symmetric case, the probability of reaching consensus on either state is equal (1/2) and remains independent of connectivity. In contrast, in the asymmetric case, the probability of 0-consensus increases with connectivity and scales with the population size (see Fig. \ref{fig3}).

Regarding the time required to achieve consensus, such as 0-consensus (where all agents adopt state 0), the behavior also differs. For the symmetric case, consensus time decreases with increasing network connectivity ($k$). However, in the asymmetric case, consensus time increases with $k$, exhibiting a minimum when both states are considered equally likely to achieve consensus (see Fig. \ref{fig4}).

This asymmetry is particularly pronounced when a stubborn agent is included in the population. If the stubborn agent holds state 0 and interacts with its nearest neighbors, the consensus time is significantly reduced. Conversely, if the stubborn agent permanently holds state 1, with a neighborhood consisting of the friends of its closest friends (excluding the latter), the time to achieve consensus in state 1 increases dramatically for low $k$-values (see Fig. \ref{fig5}).

To explain this symmetry breaking, we solve the recursive equation (Eq. \ref{recursive2}) after two time steps for the case $k = 2$, finding that the probability of a node being in state 0 exceeds one half—unlike in the symmetric case. It is important to note that this asymmetry arises from the specific contagion rule used. For example, if the update rule is modified in tie situations—when a node has an equal number of neighbors in each state—different probabilities of reaching 0-consensus emerge in the asymmetric case: (i) if the node retains its current state, the probability of reaching 0-consensus is 0.5, and (ii) if the node switches its state, the probability drops below 0.5, in contrast to the behavior of the classical voter model (see Fig.\ref{fig2}).

These results demonstrate that incorporating an asymmetric neighborhood significantly alters the opinion dynamics of the population, leading to several reflections on potential applications of this model, some of which are explored in the next section.

\section{Concluding remarks}

The model and results presented in this paper could be particularly relevant in situations where contrasting opinions from different sources determine the final outcome. A familiar example, as described in the introduction, involves the advice provided by close friends compared to that of less familiar individuals. Close friends often tend to be more polite or hesitant to offer critical feedback, whereas friends-of-friends may provide more honest, albeit less personalized, opinions. This dynamic can sometimes lead to misleading decisions.

In a broader context, this model can help understand how misinformation or highly polarized opinions-spreads through direct, peer-to-peer connections (first-degree neighbors), while fact-checking or corrective information may rely on more extended networks (second-degree neighbors or more distant sources). Misinformation tends to thrive within immediate social circles, reinforcing echo chambers, as individuals often prioritize interactions within their close-knit networks. Conversely, people seeking accurate information or corrections may need to rely on second-degree connections, such as fact-checkers or experts, who lie outside their immediate peer group. In this context, the model offers insights into the persistence of polarized views and explains why corrective information often takes longer to propagate through social networks.

On a similar note, the model could also provide valuable insights into political campaign strategies. It could be used to study how certain ideological groups primarily interact within their close communities (first-degree neighbors), while others may be more influenced by second-degree connections, such as political analysts or external media outlets. The findings from this study could guide campaigners in effectively targeting specific groups by adopting tailored outreach strategies. For example, campaigns could focus on leveraging first-degree influencers for tight-knit communities or second-degree influencers for more diffuse audiences, depending on the nature of the opinion being addressed.

\appendix
\section{The origin of the asymmetry for $k=4$}\label{apend1}

As it has been shown, the presence of different neighborhoods leads to asymmetric opinion dynamics in the voter model, influenced by the network's connectivity. The winning opinion arises from the nearest neighbors, prevailing over those generated by the second-nearest neighbors, i.e., the friends of the closest friends. This asymmetry becomes more pronounced in networks with low connectivity, particularly at the minimum connectivity $k=2$ (see section \ref{Voter}).

Similarly to the case $k=2$, we can calculate the probability of achieving any of the two states, $x=0$ (local neighborhood) and $x=1$ (nonlocal neighborhood), after two steps for $k=4$. For this connectivity, the smallest interacting set that affects the central agent state after two steps has 17 nodes, which yields $2^{17}$ possible configurations at step $n$. 

To know the state of the central node two step ahead, i.e at $x(n+2)$, the following scheme can be applied:
\begin{align}\label{recursiveA1}
x_i(n+1)    =   F( x_{i-4}(n),x_{i-3}(n),x_{i-2}(n), x_{i-1}(n),x_i(n),  x_{i+1}(n), x_{i+2}(n), x_{i+3}(n), x_{i+4}(n))
\end{align}
where $i = -4, \ldots, 4$ and the function $F$ is given by:
{\small
\begin{align}\label{app1}
& F(x_{i-4}(n),x_{i-3}(n),x_{i-2}(n), x_{i-1}(n),x_i(n), x_{i+1}(n), x_{i+2}(n), x_{i+3}(n), x_{i+4}(n))  = \notag \\
&\quad (1-x_i(n)) \notag \\
&\quad (x_{i-1}(n)(1-x_{i-1}(n))(1-x_{i-2}(n))(1-x_{i+1}(n))(1-x_{i+2}(n)) \notag \\
&\quad +   \mathcal{B}(0.25)* \notag \\
&\quad (  x_{i-1}(n)(1-x_{i-2}(n))(1-x_{i+1}(n))(1-x_{i+2}(n))+ \notag \\
&\quad x_{i-2}(n)(1-x_{i-1}(n))(1-x_{i+1}(n))(1-x_{i+2}(n))+ \notag \\
&\quad x_{i+1}(n)(1-x_{i-1}(n))(1-x_{i-2}(n))(1-x_{i+2}(n))+ \notag \\ 
&\quad x_{i+2}(n)(1-x_{i-1}(n))(1-x_{i-2}(n))(1-x_{i+1}(n)) ) \notag \\
&\quad +   \mathcal{B}(0.5)* \notag \\
&\quad (  x_{i-1}(n)x_{i-2}(n)(1-x_{i+1}(n))(1-x_{i+2}(n))+ \notag \\
&\quad x_{i-1}(n)x_{i+1}(n)(1-x_{i-2}(n))(1-x_{i+2}(n))+ \notag \\
&\quad x_{i-1}(n)x_{i+2}(n)(1-x_{i-2}(n))(1-x_{i+1}(n))+ \notag \\
&\quad x_{i-2}(n)x_{i+1}(n)(1-x_{i-1}(n))(1-x_{i+2}(n))+ \notag \\
&\quad x_{i-2}(n)x_{i+2}(n)(1-x_{i-1}(n))(1-x_{i+1}(n))+ \notag \\
&\quad x_{i+1}(n)x_{i+2}(n)(1-x_{i-1}(n))(1-x_{i-2}(n))   ) \notag \\
&\quad +   \mathcal{B}(0.75)* \notag \\
&\quad (   x_{i-1}(n)x_{i-2}(n)x_{i+1}(n)(1-x_{i+2}(n))+ \notag \\
&\quad x_{i-1}(n)x_{i-2}(n)x_{i+2}(n)(1-x_{i+1}(n))+ \notag \\
&\quad x_{i-1}(n)x_{i+1}(n)x_{i+2}(n)(1-x_{i-2}(n))+ \notag \\
&\quad x_{i-2}(n)x_{i+1}(n)x_{i+2}(n)(1-x_{i-1}(n))   ) \notag \\
&\quad +  x_{i-1}(n)x_{i-2}(n)x_{i+1}(n)x_{i+2}(n)) \notag \\
&\quad +  x_i(n) \notag \\
&\quad (x_{i-3}(n)(1-x_{i-3}(n))(1-x_{i-4}(n))(1-x_{i+3}(n))(1-x_{i+4}(n)) \notag \\
&\quad +  \mathcal{B}(0.25)* \notag \\
&\quad (   x_{i-3}(n)(1-x_{i-4}(n))(1-x_{i+3}(n))(1-x_{i+4}(n))+ \notag \\
&\quad x_{i-4}(n)(1-x_{i-3}(n))(1-x_{i+3}(n))(1-x_{i+4}(n))+ \notag \\
&\quad x_{i+3}(n)(1-x_{i-3}(n))(1-x_{i-4}(n))(1-x_{i+4}(n))+ \notag \\
&\quad x_{i+4}(n)(1-x_{i-3}(n))(1-x_{i-4}(n))(1-x_{i+3}(n))   ) \notag \\
&\quad +   \mathcal{B}(0.5)* \notag \\
&\quad (   x_{i-3}(n)x_{i-4}(n)(1-x_{i+3}(n))(1-x_{i+4}(n))+ \notag \\
&\quad x_{i-3}(n)x_{i+3}(n)(1-x_{i-4}(n))(1-x_{i+4}(n))+ \notag \\
&\quad x_{i-3}(n)x_{i+4}(n)(1-x_{i-4}(n))(1-x_{i+3}(n))+ \notag \\
&\quad x_{i-4}(n)x_{i+3}(n)(1-x_{i-3}(n))(1-x_{i+4}(n))+ \notag \\
&\quad x_{i-4}(n)x_{i+4}(n)(1-x_{i-3}(n))(1-x_{i+3}(n))+ \notag \\
&\quad x_{i+3}(n)x_{i+4}(n)(1-x_{i-3}(n))(1-x_{i-4}(n))   ) \notag \\
&\quad +   \mathcal{B}(0.75)* \notag \\
&\quad (   x_{i-3}(n)x_{i-4}(n)x_{i+3}(n)(1-x_{i+4}(n))+ \notag \\
&\quad x_{i-3}(n)x_{i-4}(n)x_{i+4}(n)(1-x_{i+3}(n))+ \notag \\
&\quad x_{i-3}(n)x_{i+3}(n)x_{i+4}(n)(1-x_{i-4}(n))+ \notag \\
&\quad x_{i-4}(n)x_{i+3}(n)x_{i+4}(n)(1-x_{i-3}(n))   ) \notag \\
&\quad +  x_{i-3}(n)x_{i-4}(n)x_{i+3}(n)x_{i+4}(n))
\end{align}

}
Here $x_{i \pm j}$ denote the states of the left and right first and second neighbors for each node $i$ and $\mathcal{B}(0.5)$ represents a Bernoulli distribution of $p=1/2$ in $\{0,1\}$. 

The state of the central node at time $n+2$ is then given by:
\begin{align}\label{recursiveA3}
 x_0(n+2)    =   F(&x_{-4}(n+1),x_{-3}(n+1),x_{-2}(n+1), x_{-1}(n+1),x_0(n+1), \notag \\
&  x_{+1}(n+1), x_{+2}(n+1), x_{+3}(n+1), x_{+4}(n+1))\nonumber
\end{align}

For each initial configuration of the $2^{17}$ possible, we can obtain the probability of the central site being in state $x=0$ after two steps. As in the case $k=2$, the results also exceeds 0.5, approximately 0.508174, although closer to symmetry than in  the case $k=2$. 

\begin{figure}[H]
\centering
\includegraphics[width=0.8\textwidth]{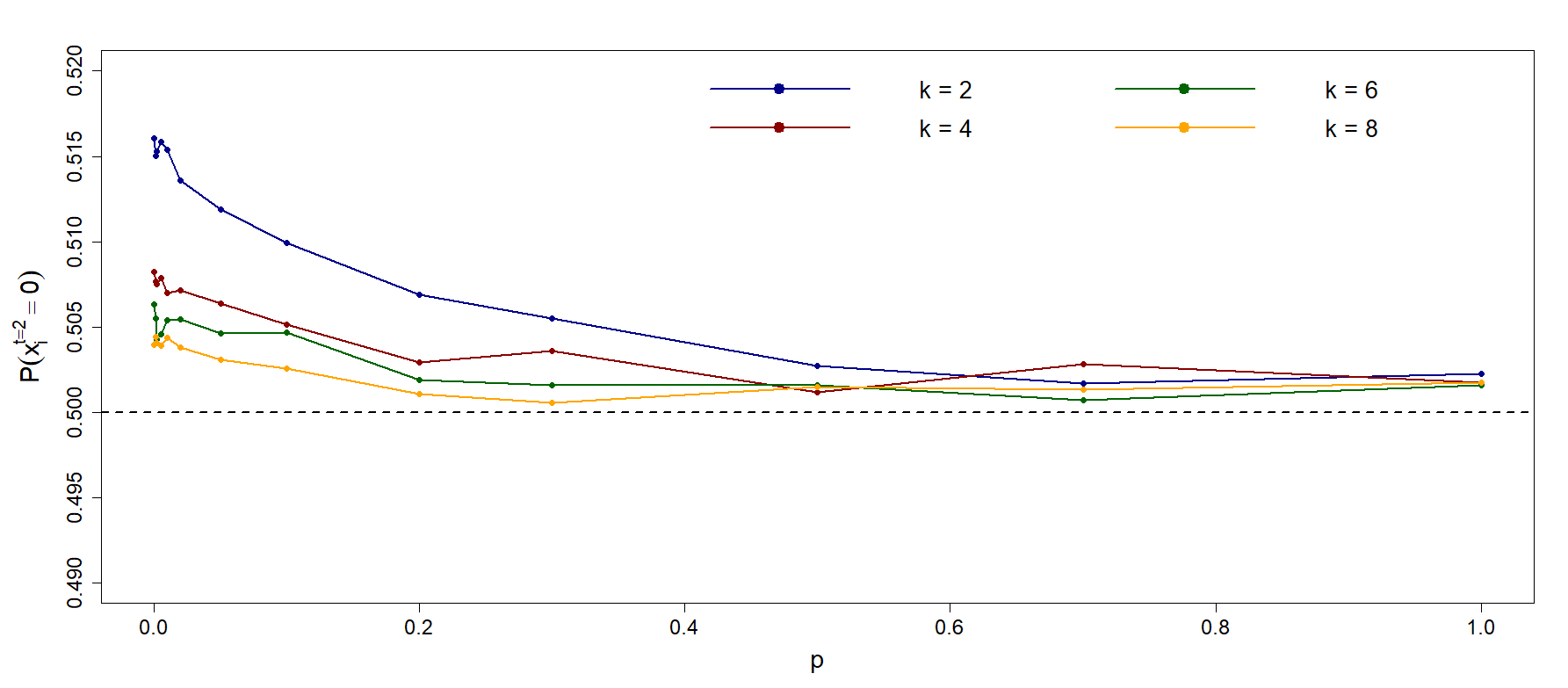}
\caption{Probability of the central site reaching state 0 after two steps as a function of the rewiring probability $p$ in the asymmetric case. The curves consider all possible initial state configurations of the influencing sites, for different values of $k$. Each point in the curves are obtained as an average from 1000 initial configurations. As observed, the probability of the central site reaching state 0 exceeds 0.5 for the four $k$ values of the connectivity. This explain why, asymptotically, the 0-state, primarily influenced by nearest neighbor, is more likely than the 1-state, which is influenced by second neighbors (see Fig. \ref{fig6}). Note that the points at $p=0$ for $k=2$ and $k=4$ provide the theoretical values stated in the main text and in this appendix, respectively.}
\label{fig8}
\end{figure}

This asymmetry is also present for positive values of $p$ (rewiring probability). Fig.\ref{fig8} shows the probability that the central site is in state 0 after two steps, starting from all possible state combinations of the influencing sites, for several values of $k$. As seen in the figure, this probability is consistently greater than 0.5, highlighting the asymmetry toward the 0-state. This bias arises from the influence of nearest neighbors, which contrasts with the second-neighbor influence on 1-state agents. This observation helps explain the results presented in Figure \ref{fig6}.

%\section*{References}

%

\end{document}